\documentclass[conference,a4paper]{IEEEtran}
\usepackage{ifpdf}

\ifCLASSINFOpdf
  \usepackage[pdftex]{graphicx}
\else
\fi
\usepackage{amsmath}
\usepackage{amsfonts}
\usepackage{amssymb}
\usepackage{amsthm}
\usepackage{algorithmic}

\usepackage{array}
\usepackage{url}

\newtheorem{theorem}{Theorem}
\newtheorem{definition}{Definition}
\newtheorem{example}{Example}

\newtheorem{corollary}{Corollary}
\newtheorem{lemma}{Lemma}
\newtheorem{remark}{Remark}

\begin{document}
%
\title{Uniformity Properties of Construction C}

\author{\IEEEauthorblockN{Maiara F. Bollauf}
\IEEEauthorblockA{Institute of Mathematics, Statistic and Computer Science \\
University of Campinas, S\~ao Paulo\\
13083-859, Brazil \\
Email: maiarabollauf@ime.unicamp.br }
\and
\IEEEauthorblockN{Ram Zamir}
\IEEEauthorblockA{Dept. Electrical Engineering-Systems\\
Tel Aviv University, Tel Aviv \\
69978, Israel \\
Email: zamir@eng.tau.ac.il }}


%


\maketitle

\begin{abstract}
	Construction C (also known as Forney's multi-level code formula) forms a Euclidean code for the additive white Gaussian noise (AWGN) channel from $L$ binary code components. If the component codes are linear, then the minimum distance is the same for all the points, although the kissing number may vary.  In fact, while in the single level ($L=1$) case it reduces to lattice Construction A, a multi-level Construction C is in general not a lattice. We show that the two-level ($L=2$) case is special: a two-level Construction C satisfies Forney's definition for a geometrically uniform constellation. Specifically, every point sees the same configuration of neighbors, up to a reflection of the coordinates in which the lower level code is equal to 1. In contrast, for three levels and up ($L\geq 3$), we construct examples where the distance spectrum varies between the points, hence the constellation is not geometrically uniform.
\end{abstract}



%
\IEEEpeerreviewmaketitle

\section{Introduction}
	Constructing lattices based on codes is an active topic of study since its first comprehensive introduction by Conway and Sloane \cite{conwaysloane}. However, not all of the proposed constructions provide a lattice and this is the case of Construction C (or multi-level Construction by a Code-Formula \cite{forney1}, \cite{forney}). 
	
	The first immediate relation of Construction C with a lattice construction happens when we consider a single level ($L=1$), and then we obtain simply the known Construction $A$ \cite{conwaysloane}. A recent work of Kositwattanarerk and Oggier \cite{kositoggier} have also explored the relation between Construction C and (lattice) Construction D. They showed that if we consider a family of nested linear binary codes $\mathcal{C}_{1} \subseteq \dots \subseteq \mathcal{C}_{L} \subseteq \mathbb{F}_{2}^{n},$ where these nested codes are closed under Schur product, then both Constructions coincide and we obtain a lattice from Contruction C. A well-known example of a lattice construction originating from Construction C (which coincides with Construction D), that uses as underlying codes the family of Reed-Muller codes, is the Barnes-Wall lattice \cite{forney1}, \cite{forney}. 
	
	Besides that, there exist significant properties and applications of Construction C that can be useful for engineering purposes, such as the fact that Construction C with multi-stage decoding can achieve the high SNR capacity of an AWGN channel asymptotically as the dimension $n$ goes to infinity \cite{forneytrottchung}. Moreover, if the underlying codes of this construction are linear, then all points in this constellation have the same minimum distance
\cite{conwaysloane}, but not necessarily the same kissing number (our following Example \ref{cex1} shows, for example, that the kissing number of an element of Construction C varies between $1$ and $2$).
 
	Another application of nonlattice construction is presented by Agrell and Eriksson [\ref{agrelleriksoon}], where they proved that the $D_{n}+$ tessellation [\ref{conwaysloane}] (which we can visualize as a $2-$level Construction C) exhibits as a lower normalized second moment (i.e. a better quantization efficiency) than any known lattice tessellation in dimensions $7$ and $9.$ Note that a tessellation of an $n-$dimensional space is a partition of $\mathbb{R}^{n}$ into regions, such that any pair of regions can be transformed into each other through a rotation, reflection or translation, so it is generally not a lattice.
	
	Motivated by this background, our idea in this paper is to study properties of a general Construction C and find out how close to a lattice can this construction be, in case it does not satisfy the condition given by \cite{kositoggier}. Our study demonstrates that a two-level ($L=2$) Construction C provides a geometrically uniform constellation [\ref{forneyguc}], however for three levels and up ($L\geq 3$) the distance spectrum varies between the points of the constellation and consequently it is not geometrically uniform.
	
	This paper is organized as follows: in Section II we briefly define and present some known properties of Construction C. In Section III we prove that we can obtain a geometrically uniform constellation by a two-level Construction C. Finally, in Section IV we discuss the equi-distance spectrum (EDS) of a two-level Construction C and present examples which demonstrate that for more than two levels this property is not valid.


 

\section{Construction C over binary codes}

\begin{definition}(\textit{Construction C})  Consider $L$ binary codes $\mathcal{C}_{1}, \dots, \mathcal{C}_{L} \subseteq \mathbb{F}_{2}^{n},$ not necessarily nested or linear. Then we define an infinite constellation $\Gamma_{C}$ in $\mathbb{R}^{n}$ that is called Construction C as:
\begin{equation} \label{eqC}
\Gamma_{C}=\mathcal{C}_{1}+2\mathcal{C}_{2}+ \dots + 2^{L-1}\mathcal{C}_{L}+2^{L}\mathbb{Z}^{n},
\end{equation} 
where $+$ denotes real addition, and $\mathbb{Z}$ denotes the set of integers.
\end{definition}

	Note that if $L=1,$ i.e., if we consider a single level with a linear code, then this construction reduces to lattice Construction A.


\begin{example} Consider $\mathcal{C}_{1}=\{(0,0),(1,1)\}$ and $\mathcal{C}_{2}=\{(0,0)\}.$ Then, applying the process defined by Construction $C$ we have:
\begin{equation}
\Gamma_{C}=\mathcal{C}_{1}+2\mathcal{C}_{2}+4\mathbb{Z}^{2}.
\end{equation}

	Geometrically, we can see this constellation in Figure \ref{ex1} where clearly $\Gamma_{C}$ is not a lattice.
	
\begin{figure}[h]
\begin{center}
		\includegraphics[height=6cm]{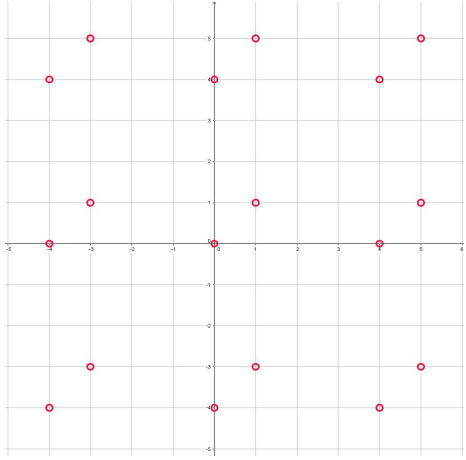}  
\caption{{$\Gamma_{C}$ obtained by Construction $C$}} \label{ex1}
\end{center}
\end{figure}
\end{example}

	
	In general, Construction $C$ produces a nonlattice constellation. There is a modification of this construction called  Construction $D,$ that always provides a lattice. This happens when the codes $\mathcal{C}_{i}$ considered in the construction are linear and nested, i.e., $\mathcal{C}_{i} \subseteq \mathcal{C}_{i+1},$ and we combine their basis vectors rather than the full codes $\mathcal{C}_{i}$ as in (\ref{eqC}).
	
\begin{definition} (\textit{Construction D})  Let $\mathcal{C}_{1} \subseteq \dots \subseteq \mathcal{C}_{L} \subseteq \mathbb{F}_{2}^{n}$ be a family of nested linear binary codes. 
Let $k_{i}=\dim(\mathcal{C}_{i})$ and let $b_{1}, b_{2}, \dots, b_{n}$ be a basis of $\mathbb{F}_{2}^{n}$ such that $b_{1}, \dots, b_{k_{i}}$ span $\mathcal{C}_{i}.$ The lattice $\Lambda_D$ consists of all vectors of the form
\begin{equation}
\displaystyle\sum_{i=1}^{L} 2^{i-1} \displaystyle\sum_{j=1}^{k_{i}} \alpha_{j}^{i} \psi(b_{j})+2^{L}z
\end{equation}
where $\alpha_{j}^{i} \in \{0,1\},$ $z \in \mathbb{Z}^{n}$ and the map $\psi: \mathbb{F}_{2}^{n} \rightarrow \mathbb{R}^{n}$ is defined by
\begin{equation}
\psi(x_{1}, \dots, x_{n})=(\psi_{i}(x_{1}), \dots, \psi_{i}(x_{n})),
\end{equation}
with $\psi_{i}(x_{k})=x_{k},$ for $i=1, \dots, L.$
\end{definition}


\begin{definition} (\textit{Schur product}) For $x=(x_{1}, \dots, x_{n})$ and $y=(y_{1}, \dots, y_{n}) \in \mathbb{F}_{2}^{n},$ we define $x \ast y = (x_{1}y_{1}, \dots, x_{n}y_{n}).$
\end{definition}

	Denote by $\Lambda_{C}$ the smallest lattice that contains $\Gamma_{C}.$ There is a condition that if satisfied can guarantee that the Construction C will provide a lattice.
	
\begin{theorem} \cite{kositoggier} (\textit{Relation between Constructions C and D}) Given a family of nested binary linear codes $\mathcal{C}_{1} \subseteq \dots \subseteq \mathcal{C}_{L} \subseteq \mathbb{F}_{2}^{n},$ then the following statements are equivalent:
\begin{itemize}[\IEEEsetlabelwidth{Z}]
\item[1.] $\Gamma_{C}$ is a lattice.
\item[2.] $\Gamma_{C}=\Lambda_{C}.$
\item[3.] $\mathcal{C}_{1} \subseteq \dots \subseteq \mathcal{C}_{L} \subseteq \mathbb{F}_{2}^{n}$ is closed under Schur product.
\item[4.] $\Gamma_{C}=\Lambda_{D},$ 
\end{itemize}
\end{theorem} 

	From our point of view, the result shows a condition under which Construction C provides a lattice constellation and this condition makes Constructions C and D coincide.
	
\begin{example} The $D_{n}+$ tessellation [\ref{conwaysloane}] could be conceived as a $2-$ level Construction C if we consider $\mathcal{C}_{1}$ as the $[n,1,n]-$ repetition code and $\mathcal{C}_{2}$ as the $[n,n-1,2]-$even parity check code. Note that for $n$ even, this construction represents a lattice, because we would have nested linear codes that are closed under Schur product. Otherwise, when $n$ is odd, we obtain a nonlattice constellation which coincides with our Construction C. In particular, for dimensions $n=7$ and $9$ it was proved that $D_{n}+$ has a lower normalized second moment than any known lattice tessellation [\ref{agrelleriksoon}]. 

\end{example}

\section{Geometric uniformity for L=2 linear levels}

	An important lattice property is geometric uniformity and the first point of this study is to check if the constellation given by $\Gamma_{C}$ satisfies this property.
	
\begin{definition} \label{gudef} (\textit{Geometrically uniform constellation}) A constellation $\Gamma$ is geometrically uniform if for any two codewords $c,c' \in \Gamma$ there exists a distance-preserving transformation $T$ (translation, reflection, rotation, permutation) such that $c'=T(c)$ and $T(\Gamma)=\Gamma.$
\end{definition}

\begin{remark} Note that the notion of geometric uniformity can be redefined such that for every point $x \in \Gamma$ there exists a transformation $T$ consisting only of reflection, rotation or permutation such that
\begin{equation} \label{eqdefgu}
T(\Gamma-x)=\Gamma.
\end{equation}
This definition is equivalent to Definition \ref{gudef} by replacing $c'=0.$
\end{remark}

\begin{example} Every lattice $\Lambda$ is geometrically uniform, due to the fact that any translation $\Lambda+x$ by a lattice point $x \in \Lambda$ is just $\Lambda$ and it means that every point of the lattice has the same number of neighbors at each distance and all Voronoi regions are congruent. Indeed, any lattice translation $\Lambda+t$ is geometrically uniform.
\end{example}

\begin{theorem} \label{propgu} (Geometric uniformity of $2-$level Construction C) Consider $\Gamma_{C}=\mathcal{C}_{1}+2\mathcal{C}_{2}+ 4\mathbb{Z}^{n}$ where $\mathcal{C}_{1}$ and $\mathcal{C}_{2}$ are linear codes. Then, $\Gamma_{C}$ is geometrically uniform.
\end{theorem}

\begin{IEEEproof} 
Let $\Gamma_{C}=\mathcal{C}_{1}+2\mathcal{C}_{2}+ 4\mathbb{Z}^{n}$ and fix $x=(x_{1},x_{2}, \dots, x_{n})$ given by
\begin{center} 
$x=c_{1}+2c_{2}+4z=(c_{11},\dots, c_{1n})+2(c_{21}, \dots, c_{2n})+ $ \\
$4(z_{1}, \dots, z_{n}),$
\end{center}
where $c_{1} \in \mathcal{C}_{1}, c_{2} \in \mathcal{C}_{2}.$ Consider \begin{equation} \label{eqT} T=\begin{pmatrix}
	(-1)^{c_{11}} & 0 & \dots & 0 \\
	0 & (-1)^{c_{12}} & \dots & 0  \\
	0 & 0 & \ddots & 0 \\
	0 & 0 & \dots & (-1)^{c_{1n}} 
	\end{pmatrix}, \end{equation} that is, T reflects the coordinates in which $c_{1}$ is equal to one and maintains the coordinates where $c_{1}$ is equal to zero. 
	
	 Let $y \in \Gamma_{C}$ be given by $y=(y_{1},y_{2}, \dots, y_{n})$ with
\begin{equation*}
y=\tilde{c_{1}}+2\tilde{c_{2}}+4\tilde{z}=(\tilde{c_{11}},\dots, \tilde{c_{1n}})+2(\tilde{c_{21}}, \dots, \tilde{c_{2n}}+
\end{equation*}
\begin{equation}
4(\tilde{z_{1}}, \dots, \tilde{z_{n}}),
\end{equation} 
where $\tilde{c_{1}} \in \mathcal{C}_{1}, \tilde{c_{2}} \in \mathcal{C}_{2}.$  Following (\ref{eqdefgu}), we first need to show that $T(y-x) \in \Gamma_{C}.$ We write
\begin{equation}
y-x=(\tilde{c_1}-c_{1})+2(\tilde{c_{2}}-c_{2}) +4(\tilde{z}-z) 
\end{equation}
and 
\begin{equation*}
T(y-x)=\begin{pmatrix}
	(-1)^{c_{11}} & 0 & \dots & 0 \\
	0 & (-1)^{c_{12}} & \dots & 0  \\
	0 & 0 & \ddots & 0 \\
	0 & 0 & \dots & (-1)^{c_{1n}} 
	\end{pmatrix} \cdot \nonumber
\end{equation*}	
\begin{equation*} \underbrace{\begin{pmatrix}
	(\tilde{c_{11}}-c_{11})+2(\tilde{c_{21}}-c_{21})+4(\tilde{z_{1}}-z_{1}) \\
	(\tilde{c_{12}}-c_{12})+2(\tilde{c_{22}}-c_{22})+4(\tilde{z_{2}}-z_{2}) \\
	\vdots \\
	(\tilde{c_{1n}}-c_{1n})+2(\tilde{c_{2n}}-c_{2n})+4(\tilde{z_{n}}-z_{n}) 
	\end{pmatrix}}_{y-x}. \nonumber
\end{equation*}

	Hence, the $i-$th coordinate of $T(y-x)$ has two possible cases:
{\small \begin{equation} \label{exT}
 [T(y-x)]_{i}= \begin{cases} 
   \tilde{c_{1i}}+2(\tilde{c_{2i}}-c_{2i})+4(\tilde{z_{i}}-z_{i}), & \text{if } c_{1i}=0 \\
   (1-\tilde{c_{1i}})- 2(\tilde{c_{2i}}-c_{2i})-4(\tilde{z_{i}}-z_{i}), & \text{if } c_{1i}=1
\end{cases}
\end{equation} }
for $i=1,\dots,n.$

	We can rewrite the expression (\ref{exT}) as
\begin{equation*}
 [T(y-x)]_{i}= (\tilde{c_{1i}}-c_{1i}) {\rm mod} \ 2 \ + 2[(\tilde{c_{2i}}-c_{2i}) {\rm mod} \ 2] + 4z_{i}',
\end{equation*}
where
\begin{equation}
z_{i}'= \begin{cases} 
   \tilde{z_{i}}-z_{i}, & \text{if } c_{1i}=0 \text{ and } \tilde{c_{2i}}-c_{2i} \geq 0 \\
   \tilde{z_{i}}-z_{i}-1, & \text{if } c_{1i}=0 \text{ and } \tilde{c_{2i}}-c_{2i} < 0 \\
   z_{i}-\tilde{z_{i}}, & \text{if } c_{1i}=1 \text{ and } \tilde{c_{2i}}-c_{2i} \leq 0 \\
   z_{i}-\tilde{z_{i}}-1, & \text{if } c_{1i}=1 \text{ and } \tilde{c_{2i}}-c_{2i} > 0.   
\end{cases}
\end{equation}

	This covers all the possibilities which guarantees that $T(y-x)$ is an element of $\Gamma_{C}.$ To prove that  $T(\Gamma_C -x) = \Gamma_C,$ we still need to show the reverse statement, i.e., that for each $y \in \Gamma_{C}$ there exists $y' \in \Gamma_{C}$ such that $T(y'-x)=y.$ However, this fact follows from the above derivation because $T(y-x)$ is an isometry (as a function of $y$). Therefore, we can conclude that for $L=2,$ $\Gamma_{C}$ is geometrically uniform. 

\end{IEEEproof}

\begin{remark} Observe that the choice of the transformation T in Equation (\ref{eqT}) is only with respect to the lower level code $\mathcal{C}_{1}$ due to the fact that if it was the zero code, we relapse into Construction A (inflated by a factor of $2)$ that always provides a lattice which is geometrically uniform. Therefore, this property only can be changed by the action of the code $\mathcal{C}_{1}.$
\end{remark}
	
\section{Equi-distance spectrum} 

	Geometric uniformity implies, in particular, that all points have the same set of Euclidean distances to their neighbors.

\begin{definition} (\textit{Distance spectrum}) The distance spectrum $N(d),$ for $d \geq d_{min}$ is defined as 
\begin{center}
$N(d)=$ number of points in the constellation at a Euclidean distance $d$ from the origin.
\end{center}

	We can also define
\begin{center}
$N(c,d)=$ number of points in the constellation at a Euclidean distance $d$ from an element $c$ in the constellation.
\end{center}
\end{definition}

\begin{definition} (\textit{Equi-distance spectrum}) A constellation $\Gamma$ is said to have equi-distance spectrum (EDS) if 
\begin{center}
$N(c,d)=N(d),$
\end{center}
for all $c \in \Gamma.$
\end{definition}

	Geometric uniformity implies equi-distance spectrum for $L=2$ levels in Construction C. However for $L>2$ the equi-distance spectrum (and hence the geometric uniformity) property does not hold. We will now exhibit a direct proof for this consequent property, starting by a definition of a slightly more restrictive notation of distance:

\begin{definition} (\textit{Coordinate-wise equi-distant vectors}) Two vectors are said to be coordinate-wise equi-distant if their components are equal up to a possible sign change, i.e.,
\begin{equation}
(a_{1}, \dots, a_{n}) \doteq (b_{1}, \dots,b_{n}), \ \mbox{if} \ |a_{i}|=|b_{i}|, i=1, \dots, n.
\end{equation}
\end{definition}

	We shall first show that a $2-$level Construction C satisfies the EDS property. Although this fact is clearly implied by Theorem \ref{propgu}, it will serve to contrast the counter example for three levels and up, which we shall bring next.
 
\begin{lemma} \label{lemmaequi} (\textit{Coordinate-wise equi-distant error vector in $\Gamma_{C}$})  Consider $\Gamma_{C}=\mathcal{C}_{1}+2\mathcal{C}_{2} + 4\mathbb{Z}^{n},$ where $\mathcal{C}_{1}$ and $\mathcal{C}_{2}$ are linear codes. Then, for any $x,y,x' \in \Gamma_{C},$ there exists $y' \in \Gamma_{C}$ such that the error vectors $y'-x'$ and $y-x$ are coordinate-wise equi-distant.
\end{lemma}

\begin{IEEEproof} We have that $x,y,x' \in \Gamma_{C},$ so we can write them as:
\begin{eqnarray}
x=c_1+2c_{2}+4z, \\
y=\tilde{c_1}+2\tilde{c_{2}}+4\tilde{z}, \\
x'={c_1}'+2{c_{2}}'+4{z}',
\end{eqnarray}
where $c_{1}, \tilde{c_{1}}, {c_{1}}' \in \mathcal{C}_{1}, c_{2}, \tilde{c_{2}}, {c_{2}}' \in \mathcal{C}_{2}$ and $z, \tilde{z}, z' \in \mathbb{Z}^{n}.$

	Set an element $y''= {c_{1}}''+2{c_{2}}'',$ where ${c_{1}}''= (\tilde{c_1}-c_{1}+{c_1}') \ {\rm mod} \ 2$ and ${c_{2}}''= (\tilde{c_2}-c_{2}+{c_2}') \ {\rm mod} \ 2.$ Note that, due to the linearity of the codes, clearly $y'' \in \mathcal{C}_{1}+2\mathcal{C}_{2}.$ Also define $y'=y''+4\overline{z},$ where $\overline{z} \in \mathbb{Z}^{n}$ is to be defined later and observe that $y' \in \Gamma_{C}.$
		

	By the definition of the ${\rm mod} \ 2$ operation,  for each coordinate $i = 1,\dots,n$ we have that
\begin{equation*}
|{c_{1_{i}}}''-{c_{1_{i}}}'|=|\tilde{c_{1_{i}}}-{c_{1_{i}}}| \ \mbox{and} \ |{c_{2_{i}}}''-{c_{2_{i}}}'|=|\tilde{c_{2_{i}}}-{c_{2_{i}}}|.
\end{equation*}
 
	Or equivalently, we have that either ${c_{j_{i}}}''-{c_{j_{i}}}'=\tilde{c_{j_{i}}}-{c_{j_{i}}}$ or ${c_{j_{i}}}''-{c_{j_{i}}}'=-\tilde{c_{j_{i}}}+{c_{j_{i}}},$ for $j=1,2.$ We want to show that the only degree of freedom that is available, i.e. the value of $\overline{z} \in \mathbb{Z}^{n},$ is enough to correct the possibly wrong signs that may appear when we are comparing the absolute values of $y-x$ with $y'-x'$. Hence, we want to define $\overline{z}$ such that the error vectors $y'-x'$ and $y-x$ are coordinate-wise equi-distant, i.e., $y'-x' \doteq y-x.$ 
	
	Define the following error vectors for $y-x$ 
\begin{eqnarray}
e_{1}=\tilde{c_{1}}-c_{1} \\
e_{2}=\tilde{c_{2}}-c_{2},
\end{eqnarray}
and also the error vectors for $y'-x'$
\begin{eqnarray}
e_{1}'=c_{1}''-c_{1}' \\
e_{2}'=c_{2}''-c_{2}'.
\end{eqnarray}
Clearly, $e_{1i},e_{2i}, e_{1i}',e_{2i}' \in \{-1,0,1\}$ for $i= 1,\dots,n.$

	Let $\overline{z}_{i}=z_{i}'-\tilde{z}_{i}+z_{i}+\Delta_{i},$ where $\Delta_{i} \in \{-1,0,1\}.$ We can describe the choice of $\overline{z}_{i}$ depending on the value of $\Delta_{i},$ which can be classified into the following possibilities:
\begin{itemize}[\IEEEsetlabelwidth{Z}]
\item[i)] If either $e_{1i}=0$ or $e_{2i}=0,$ then $\Delta_{i}=0.$
\item[ii)] If $|e_{1i}|=|e_{2i}|=1$ and ${\rm sign}(e_{1i})\ast{\rm sign}(e_{2i})={\rm sign}(e_{1i}')\ast {\rm sign}(e_{2i}'),$ then $\Delta_{i}=0.$
\item[iii)] If $|e_{1i}|=|e_{2i}|=1, \ {\rm sign}(e_{1i})\ast{\rm sign}(e_{2i}) \neq {\rm sign}(e_{1i}')\ast {\rm sign}(e_{2i}')$ and $e_{1i}'=e_{2i}',$ then $\Delta_{i}=-e_{1i}'.$
\item[iv)] If $|e_{1i}|=|e_{2i}|=1, \ {\rm sign}(e_{1i})\ast{\rm sign}(e_{2i}) \neq {\rm sign}(e_{1i}')\ast {\rm sign}(e_{2i}')$ and $e_{1i}' \neq e_{2i}',$ then $\Delta_{i}=e_{1i}'.$
\end{itemize}


	Proceeding this way for each coordinate we would find the element $\overline{z_{i}}$ that makes $y_{i}'=y_{i}''+4\overline{z_{i}}$ be the right candidate and then $y'=(y'_{1},y'_{2}, \dots, y'_{n})$ will be the element in $\Gamma_{C}$ which we were looking for.
\end{IEEEproof}

	In view of Lemma \ref{lemmaequi}, we have two important results that follow from it.

\begin{corollary} (\textit{Number of coordinate-wise equi-distant vectors in $\Gamma_{C}$}) For $x,x' \in \Gamma_{C}$ and $e,$ the number of elements $y'$ for which $y'-x'\doteq e$ is equal to the number of elements $y$ such that $y-x \doteq e.$ 
\end{corollary}
\begin{IEEEproof} 
 This result follows immediately from the proof of Lemma \ref{lemmaequi}, because the number of combinations that we can take to define the vector $y'$ will be the same as the number of possible combinations to define $y$ such that $y'-x'\doteq e \doteq y-x.$  
\end{IEEEproof}

	Since the Euclidean distance is a function of the coordinate distances (though not vice versa), we also have the following result:

\begin{corollary} (\textit{Equi-distance spectrum of $\Gamma_{C}$}) The distance spectrum is identical for all codewords in $\Gamma_{C},$ i.e., $N(c,d)=N(d)$ for all $c \in \Gamma_{C}.$
\end{corollary}

	The next examples will show that the result of Lemma \ref{lemmaequi} is not valid for three levels and up ($L \geq 3$).
	
\begin{example} \label{cex1} Consider the following linear codes, with $n=1$ and $L=3$:
\begin{equation*}
\mathcal{C}_{1}=\{0,1\}, \ \ \mathcal{C}_{2}=\{0,1\}, \ \ \ \mathcal{C}_{3}=\{0\}.
\end{equation*}

	The Construction C in this case will be given by $\Gamma_{C}=\mathcal{C}_{1}+2\mathcal{C}_{2}+ 4\mathcal{C}_{3}+8\mathbb{Z}.$ So, consider elements $x,y,x' \in \Gamma_{C}$ such that 
\begin{center}
$x=0+ 2 \cdot 0+4 \cdot 0+8 \cdot 0= 0,$ \\
$y=1+ 2 \cdot 1+4 \cdot 0+8 \cdot 0= 3,$ \\
$x'=1+ 2 \cdot 0+4 \cdot 0+8 \cdot 1= 9. $
\end{center}

	We want to show that there is no $y' \in \Gamma_{C}$ such that $y-x \doteq y'-x',$ which means in this one dimensional case:
\begin{equation}
|y-x|=|y'-x'|.
\end{equation}

	Note that $|y-x|=3$ and the only two values that $y'$ can take to satisfy $|y'-9|=3$ are $y'=6$ or $y'=12.$ However, the numbers that can be obtained by combination of codewords plus multiples of $8$ are
$$
0,1,2,3,8,9,10,11,16,17,18,19,24,25,\dots
$$
which do not include $6$ and $12.$

	Therefore, it is not possible to always find an element $y' \in \Gamma_{C}$ such that $y'-x' \doteq y-x,$ with $x,y,x' \in \Gamma_{C}$ and it proves that the result of Lemma \ref{lemmaequi} is not valid for three levels.
\end{example}

	Clearly, this counter example can be extended to any number of levels $L \geq 3.$

	Example \ref{cex1} relies on $\mathcal{C}_{3}$ being more sparse than $\mathcal{C}_{1}$ and $\mathcal{C}_{2}$.  However, efficient multi-level constructions usually exhibit the opposite property.  For example, Construction D requires a forward nesting relation $\mathcal{C}_1 \subseteq \mathcal{C}_{2} \subseteq \mathcal{C}_{3}$.  The next example demonstrates that even when the codes are nested (specifically, all three codes are equal), the result of Lemma \ref{lemmaequi} does not always hold for a $3-$level Construction C.
	
\begin{example} \label{ex5}
		Consider an $(n=3,L=3)$ Construction C with the following three identical component linear codes: 
\begin{center}
$\mathcal{C}_{1}=\mathcal{C}_{2}=\mathcal{C}_{3}=\{(0,0,0),(1,0,1),(1,1,0),(0,1,1)\}.$ \\
\end{center}

	Notice that $\mathcal{C}_{1} \subseteq \mathcal{C}_{2} \subseteq \mathcal{C}_{3} \subseteq \mathbb{F}_{2}^{3}$ and more than that, this chain is not closed under Schur product. Indeed, consider $x=(1,0,1)$ and $y=(1,1,0) \in \mathcal{C}_{1}$ and we have that $x \ast y= (1,0,0) \notin \mathcal{C}_{2},\mathcal{C}_{3}$ and the same elements can be used to prove that $\mathcal{C}_{2}$ is not closed under Schur product.
	
	Suppose that we had chosen the following elements in $\Gamma_{C}=\mathcal{C}_{1}+2\mathcal{C}_{2}+4\mathcal{C}_{3}+8\mathbb{Z}^{3}:$
\begin{center}
$x=(1,0,1)+2(0,0,0)+4(0,0,0)+8(0,0,0)=(1,0,1),$ \\
$y=(1,1,0)+2(0,1,1)+4(1,0,1)+8(0,0,0)=(5,3,6),$ \\
$x'=(1,1,0)+2(1,0,1)+4(0,1,1)+8(0,0,0)=(3,5,6).$ \\
\end{center}

	To check whether this constellation has equi-distance spectrum (EDS) we shall first try to find an element $y' \in \Gamma_{C}$ such that $y'-x' \doteq y-x,$ so by definition we need to have
\begin{equation}
| y_{i}'-x_{i}'|=|y_{i}-x_{i}|,
\end{equation}
for all $i=1,2,3.$\\

	By analyzing each particular case:
\begin{itemize}
\item[i=1)] $y_{1}'-3=\pm 4 \Rightarrow y_{1}'=7 \ \mbox{or} \ y_{1}'=-1;$ \\
\item[i=2)] $y_{2}'-5=\pm 3  \Rightarrow y_{2}'=8 \ \mbox{or} \ y_{2}'=2;$ \\
\item[i=3)] $y_{3}'-6=\pm 5 \Rightarrow y_{3}'=11 \ \mbox{or} \ y_{3}'=1;$
\end{itemize} 	

	We will start by investigating the first coordinate of $y'.$ To have $y_{1}'=7,$ we need to have exclusively 
\begin{equation}
y'=\underbrace{( 1 , \ , \ )}_{c_{1} \in \mathcal{C}_{1}} + 2\underbrace{( 1 , \ , \ )}_{c_{2} \in \mathcal{C}_{2}}+ 4\underbrace{( 1 , \ , \ )}_{c_{3} \in \mathcal{C}_{3}}+8\underbrace{(0 , \ , \ )}_{z \in \mathbb{Z}^{3}}.
\end{equation}
Following this process, to have $y_{2}'=8:$
\begin{equation} \label{e1}
y'=\underbrace{( 1 , 0 , \ )}_{c_{1} \in \mathcal{C}_{1}} + 2\underbrace{( 1 , 0 , \ )}_{c_{2} \in \mathcal{C}_{2}}+ 4\underbrace{( 1 , 0 , \ )}_{c_{3} \in \mathcal{C}_{3}}+8\underbrace{(0 , 1 , \ )}_{z \in \mathbb{Z}^{3}}
\end{equation} 
so the only options for $c_{1}, c_{2}$ and $c_{3}$ are respectively $(1,0,1), (1,0,1)$ and $(1,0,1).$ Observe that we cannot obtain with these elements $y_{3}'=11$ or $1,$ because $y_{3}'=7+8k,$ $k \in \mathbb{Z}.$

	If we change the value of $y_{2}'$ to $2,$ that is the other available option, we would have instead of Equation (\ref{e1}):
\begin{equation}
y'=\underbrace{( 1 , 0, \ )}_{c_{1} \in \mathcal{C}_{1}} + 2\underbrace{( 1 , 1, \ )}_{c_{2} \in \mathcal{C}_{2}}+ 4\underbrace{( 1 , 0 , \ )}_{c_{3} \in \mathcal{C}_{3}}+8\underbrace{(0 , 0 , \ )}_{z \in \mathbb{Z}^{3}},
\end{equation}
and the possible elements $c_{1}, c_{2}$ and $c_{3}$ are in this case, respectively $(1,0,1), (1,1,0)$ and $(1,0,1).$ Notice that $y_{3}'=5+8k,$ $k \in \mathbb{Z}$ and there is no integer $k$ that make the result $y_{3}'=11$ or $1$ holds.

	Notice that $7 \equiv -1(\mod \ 8)$ so taking the option of $y_{1}'=-1$ would not affect anything in the arrangement of the coordinates. In consequence of that, we can conclude that there is no $y' \in \Gamma_{C}$ such that $y'-x'$ is coordinate-wise equi-distant to $y-x.$
\end{example}

\begin{remark} This conclusion however does not imply that there is no $y'$ in $\Gamma_C$ with the same Euclidean distance  $||y'-x'|| = ||y-x|| = \sqrt{4^2+3^2+5^2} = \sqrt{50}.$ In particular, any permutation of the coordinate distances $(4,3,5)$ preserves the Euclidean distance. Indeed, the point $y' = (8,9,9) \in \Gamma_C$ satisfies $y'-x' = (5,4,3)$, so its Euclidean distance from $x'$ is the same as that of $y$ from $x.$
\end{remark}

\section{Conclusion}

	We have seen that a general Construction C has two important properties for $L=2$ levels, which are the geometric uniformity and consequently equi-distant spectrum. Moreover, we also proved that these two properties are exclusive for $L=1$ and $2$ levels, the first because it falls into the case of Construction $A$ that is a lattice construction, and the last is the main result of this work, together with the counter examples which show that the respective two properties are not always valid for $L>2$ levels.
	
	We also observed that the importance of this study is well represented by its application into quantization process, where a $2-$ level Construction C represents a low normalized second moment in $7$ and $9$ dimensions [\ref{agrelleriksoon}]. 
	
	The recent years saw the blooming of multi-terminal lattice coding theory, e.g., for side information problems, network coding and interference alignment [\ref{nazergastpar}, \ref{pradhanramchandran}, \ref{zamirshamaierez}]. Some of the new coding schemes are motivated by the insight provided by the lattice structure, while others really hinge upon the Euclidean-space linearity of the lattice.  More specifically, multi-terminal codes usually use a nested pair of codes, where one of the codes should be closed under real addition (i.e., a lattice) while the other code can have a non-linear structure.As multi-level codes are natural candidates for nesting, it would be interesting to explore the potential of a hybrid Construction C/D nested coding scheme for efficient multi-terminal coding. 

	Another interesting future work is related with Forney's research [\ref{forneyguc}], where he considered generalized coset codes in $\mathbb{R}^{n}$
based on nested geometrically uniform codes $S/S'$ and found conditions under which the resulting coset code is geometrically uniform. For the special case $S=\mathbb{Z}$ and $S'=2\mathbb{Z},$ this coset code reduces to a Construction A lattice.  Thus, it would be interesting to extend the main result of this paper
to a $2-$level generalized coset code based on a chain $S/S'/S''$ of three geometrically uniform codes.




\section*{Acknowledgment}

The first author would like to thank Sueli I. R. Costa for making this collaboration happen and also thank Eleon\'esio Strey for a meaningful dialogue around the relation between Constructions C and D. The second author would like to thank Uri Erez, Or Ordentlich and Yaron Shany for fruitful discussions regarding the EDS property of Construction C. The authors are also grateful to Erik Agrell for his thoughtful suggestions.



%

\end{document}